\documentstyle[mprocl,psfig]{article}
\makeatletter
\def\thebibliography#1{{\section*{\refname}\@mkboth
  {\refname}{\refname}}\footnotesize
  \typeout{\refname}\def\bibwidthlabel{#1}\list
  {\kapbib@counter}{\kapbib@list}
    \let\makelabel\@biblabel
    \def\newblock{\hskip .11em plus .33em minus .07em}
    \sloppy\clubpenalty4000\widowpenalty4000
    \sfcode`\.=1000\relax}

  \def\kapbib@counter{\relax}
  \def\kapbib@list{\setlength{\labelsep}{0em}%
        \setlength{\labelwidth}{0pt}%
        \setlength{\itemindent}{-\bibhang}%
        \setlength{\itemsep}{0pt}%
        \setlength{\parsep}{0pt}%
        \setlength{\leftmargin}{\bibhang}}
  \newlength{\bibhang}
  \def\@biblabel#1{}

\let\@internalcite\cite
\def\cite{\@ifstar{\citetext}{\citefull}}      
\def\citetext{\def\@citeseppen{-1000}
    \def\@cite##1##2{{##1\if@tempswa , ##2\fi}}
    \def\citeauthoryear##1##2{\rm ##1  (##2)}\@internalcite}
\def\citefull{\def\@citeseppen{-1000}%
    \def\@cite##1##2{({##1\if@tempswa , ##2\fi})}%
    \def\citeauthoryear##1##2{\rm ##1  ##2}\@internalcite}
\def\shortcite{\def\@citeseppen{1000}%
    \def\@cite##1##2{({##1\if@tempswa , ##2\fi})}%
    \def\citeauthoryear##1##2{\rm ##2}\@internalcite}
\def\citeauthor#1{\def\@citeseppen{1000}%
    \def\@cite##1##2{{##1\if@tempswa , ##2\fi}}%
    \def\citeauthoryear##1##2{\rm ##1}\@citedata{#1}}
\def\citeyear#1{\def\@citeseppen{1000}%
    \def\@cite##1##2{{##1\if@tempswa , ##2\fi}}%
    \def\citeauthoryear##1##2{\rm ##2}\@citedata{#1}}

\def\@citedata#1{\@tempswafalse%
 \if@filesw\immediate\write\@auxout{\string\citation{#1}}\fi
  \def\@citea{}\@cite{\@for\@citeb:=#1\do
    {\@citea\def\@citea{,\penalty\@citeseppen\ }\@ifundefined
       {b@\@citeb}{{\bf ?}\@warning
       {Citation `\@citeb' on page \thepage \space undefined}}%
{\csname b@\@citeb\endcsname}}}{}}

\def\@citex[#1]#2{\if@filesw\immediate\write\@auxout{\string\citation{#2}}\fi
  \def\@citea{}\@cite{\@for\@citeb:=#2\do
    {\@citea\def\@citea{;\penalty\@citeseppen\ }\@ifundefined
       {b@\@citeb}{{\bf ?}\@warning
       {Citation `\@citeb' on page \thepage \space undefined}}%
{\csname b@\@citeb\endcsname}}}{#1}}
\makeatother
\makeatletter
\def\mpspacing#1{\def\baselinestretch{#1}\let\glb@currsize=\relax\selectfont}
\renewcommand{\fps@figure}{thbp}
\renewcommand{\fps@table}{thbp}

\makeatother
\def\floatwidth{0.7\textwidth}

\def\mag{\nobreak\mbox{$^{\rm m}$}\!\!\!\!.\;}

\def\kms{\nobreak\mbox{$\;$km\,s$^{-1}$}}

\def\etal{{\em et~al.~}}
\begin{document}
%
\title{VARIATIONS OF THE COSMIC EXPANSION FIELD AND \\
           THE VALUE OF THE HUBBLE CONSTANT}

\author{G.A. TAMMANN}

\address{Astronomisches Institut der Universit\"at Basel, \\
       Venusstr.~7, CH-4102 Binningen, Switzerland}
\maketitle

\abstracts{%
Four Hubble diagrams are combined to test for the linearity of the
cosmic expansion field. The expansion rate, $H_0$, is found to
decrease by $\sim5\%$ out to $18\,000\kms$. Beyond this distance the
mean value of $H_0$ is close to the value at $10\,000\kms$. The
absolute value of $H_0$ is derived in two different ways. The one
leads through Cepheids from {\sl HST\/} and blue SNe\,Ia to $H_0=57\pm7$
(external error) at $30\,000\kms$. The other uses various distance
indicators to the Virgo cluster whose distance can be extended to
$10\,000\kms$ by means of only relative cluster distances; the result
of $H_0$ is closely the same. Independent distances from purely
physical methods (SZ effect, gravitational lenses, and MBW
fluctuations) cluster about $H_0\;$(cosmic)$\approx58$.
}

\section{Introduction}
\label{sec:1}
At small scales the cosmic expansion field is highly disturbed. The
negative recession velocity of the Andromeda nebula (M\,31) is telling
proof, as well as the local peculiar motion of $630\kms$ relative to
the MWB comprising a volume of radius $2000\kms$ at least.
The question therefore arises out to what distances one has to go that
the observed expansion rate, i.\,e.\ the Hubble constant $H_0$, has
truly cosmic significance. This problem is addressed in
Section~2, where four Hubble diagrams are combined to test
for the {\em linearity\/} of the expansion field in function of
distance. In Section~3 distance determinations via Cepheid distances
from Hubble Space Telescope ({\sl HST\/}) and blue SNe\,Ia are pushed out
to $30\,000\kms$. The distance of the Virgo cluster is derived in
Section~4 using different methods, and this distance is transported to
$10\,000\kms$ by means of relative cluster distances. Purely physical
determinations of $H_0$ are compiled in Section~5. The conclusions are
given in Section~6.

\section{The linearity of the Cosmic Expansion Field}
\label{sec:2}
The fundamental test for the linearity of the cosmic expansion field
is provided by the Hubble diagram. In its classical form it is a
diagram with the logarithm of the redshift ($\log cz$) plotted against
the apparent magnitude of standard candles. Useful standard candles
are very luminous objects whose absolute magnitude scatter is $\le
0\mag3$. The case of linear expansion requires a slope of $0.2$. Any
deviations from this slope translate directly into deviations from
linearity.

   The power of the Hubble diagram is twofold. It uses only directly
observable quantities and the resulting scatter about the mean Hubble
line is an upper limit of the luminosity scatter of the objects under
consideration and therefore tests the basic assumption of standard
candles.
\def\floatwidth{0.665\textwidth} 
\begin{figure}[t]
\centerline{\psfig{file=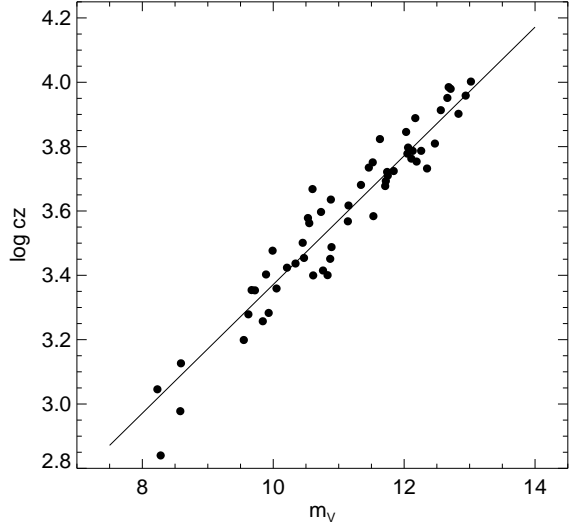,width=\floatwidth}}
\caption{The Hubble diagram of first-ranked E galaxies in nearby
  groups and clusters. Data from Sandage (1975).}
  \label{fig:1}
\end{figure}
\nocite{Sandage:75}
\begin{figure}[t]
\centerline{\psfig{file=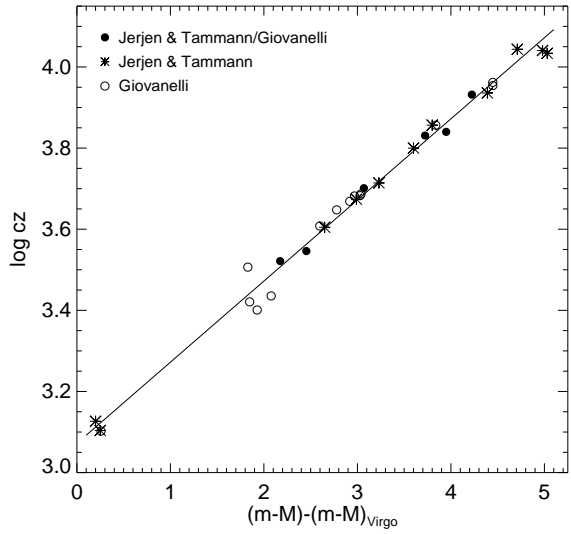,width=\floatwidth}}
\caption{Hubble diagram of 31 clusters with known relative
     distances. Asterisks are data from Jerjen \& Tammann (1993). Open
     circles are from Giovanelli (1997). Filled circles are the
     average of data from both sources.}
  \label{fig:2}
\end{figure}
\nocite{Giovanelli:97a}
\nocite{Jerjen:Tammann:93}
\begin{figure}[t]
\centerline{\psfig{file=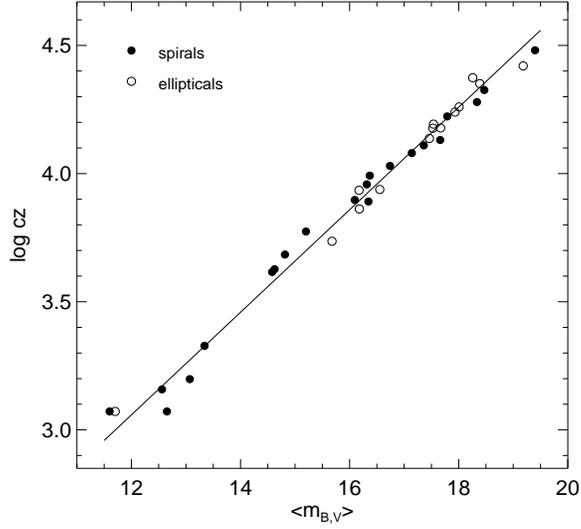,width=\floatwidth}}
\caption{The Hubble diagram of 35 blue SNe\,Ia with photometry after
  1985.\ Mean $B$ and $V$ magnitudes are plotted.\ The SNe\,Ia in
  elliptical galaxies are systematically fainter
  by $0\mag18$ than those in spiral galaxies; they are here shifted by
  this amount.\ Data mainly from Hamuy \etal (1996) with additions as
  in Saha \etal (1997) and slightly revised by
  Parodi \& Tammann (1998).}
  \label{fig:3}
\end{figure}
\nocite{Hamuy:etal:96}
\nocite{Saha:etal:97}
\nocite{Parodi:Tammann:98}
\begin{figure}[t]
\centerline{\psfig{file=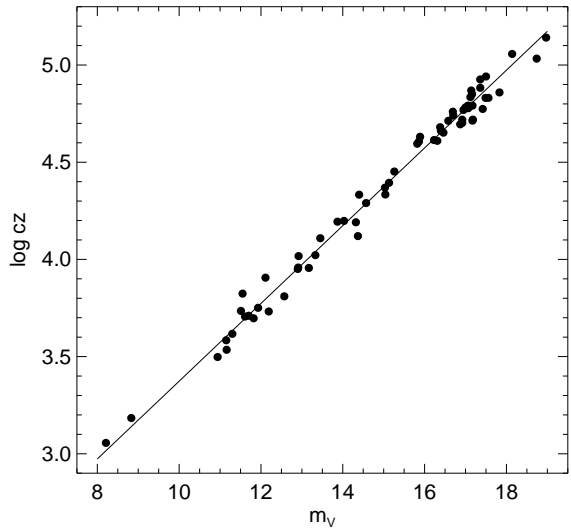,width=\floatwidth}}
\caption{The Hubble diagram of first-ranked cluster galaxies in rich
  clusters. Data from Sandage, Kristian, \& Westphal (1976).}
  \label{fig:4}
\end{figure}
\nocite{Sandage:etal:76}

   Four Hubble diagrams which reach to progressively larger redshifts
are shown in Fig.~\ref{fig:1}\,-\,\ref{fig:4}.
The nature of the Hubble diagram in Fig.~\ref{fig:2} is somewhat
different, using {\em relative\/} distance moduli of clusters instead
of standard candles. These relative moduli are much more secure than
absolute distances which are always very sensitive to sample
incompleteness and selection biases. To allow for the
different absolute magnitudes of the objects in these diagrams
(and for the use of relative cluster moduli in Fig.~\ref{fig:2}) they
are combined into a single Hubble diagram in Fig.~\ref{fig:5} by a
corresponding shift along the abscissa. The data are well fit by
a Hubble line of slope 0.2.
A free fit gives a slope of $0.197\pm0.003$, which translates into
$\log cz=H_{0} r^{0.985\pm0.015}$ (cf. \citeauthor{Sandage:etal:72}
1972; \citeauthor{Jerjen:Tammann:93} 1993). The nearly perfectly
linear ex\-pan\-sion is therefore demonstrated over a very large velocity
interval.

\def\floatwidth{0.8\textwidth} 
\begin{figure}[b]
  \centerline{\psfig{file=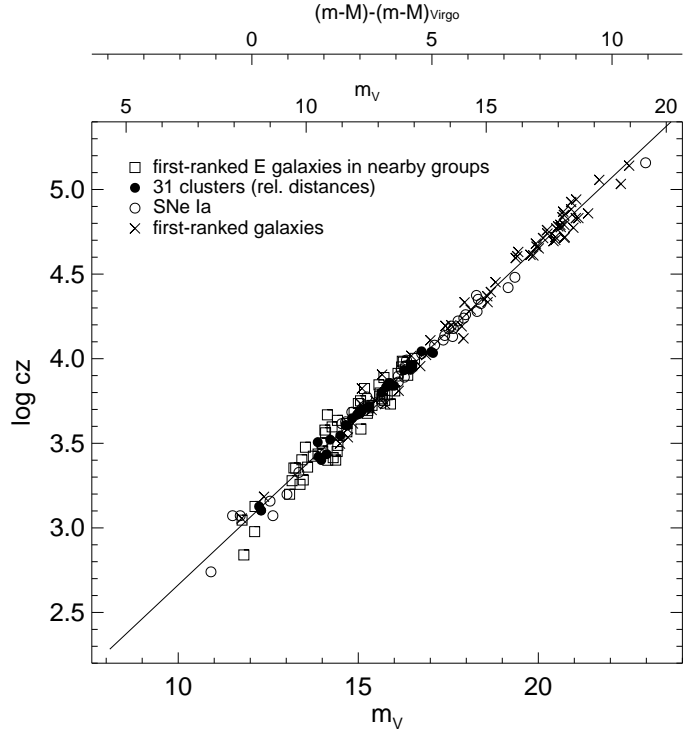,width=\floatwidth}}
  \caption{Combined Hubble diagram from Fig. 1-4.}
  \label{fig:5}
\end{figure}

   The next step is to test for {\em local\/} deviations from the
linear flow. To do this
the relative peculiar motions $\Delta v/v_{\rm c}$ are
plotted against the distance $r$ (or in sufficient approximation
against the recession velocity $v$).
If $H_{i}=(v_{\rm c} + \Delta v)/{r}$ is the perturbed
value of the Hubble ratio of the $i$-th object at distance
$r$, and $H_{0}= <\!\!H_{i}\!\!> ={v_{\rm c}}/{r}$ the true Hubble
constant, than ${\Delta H}/{H_0}=({H_{i}-H_{0}})/{H_0}=
{\Delta v}/{v_{\rm c}}$.
In Fig.~\ref{fig:5} all residuals ${\Delta v}/{v_{\rm c}}$
are read and combined within $5000\kms$ bins.
Sliding means in $2500\kms$ steps are plotted in Fig.~\ref{fig:6}
against redshift.
\def\floatwidth{0.75\textwidth}
\begin{figure}[t]
  \centerline{\psfig{file=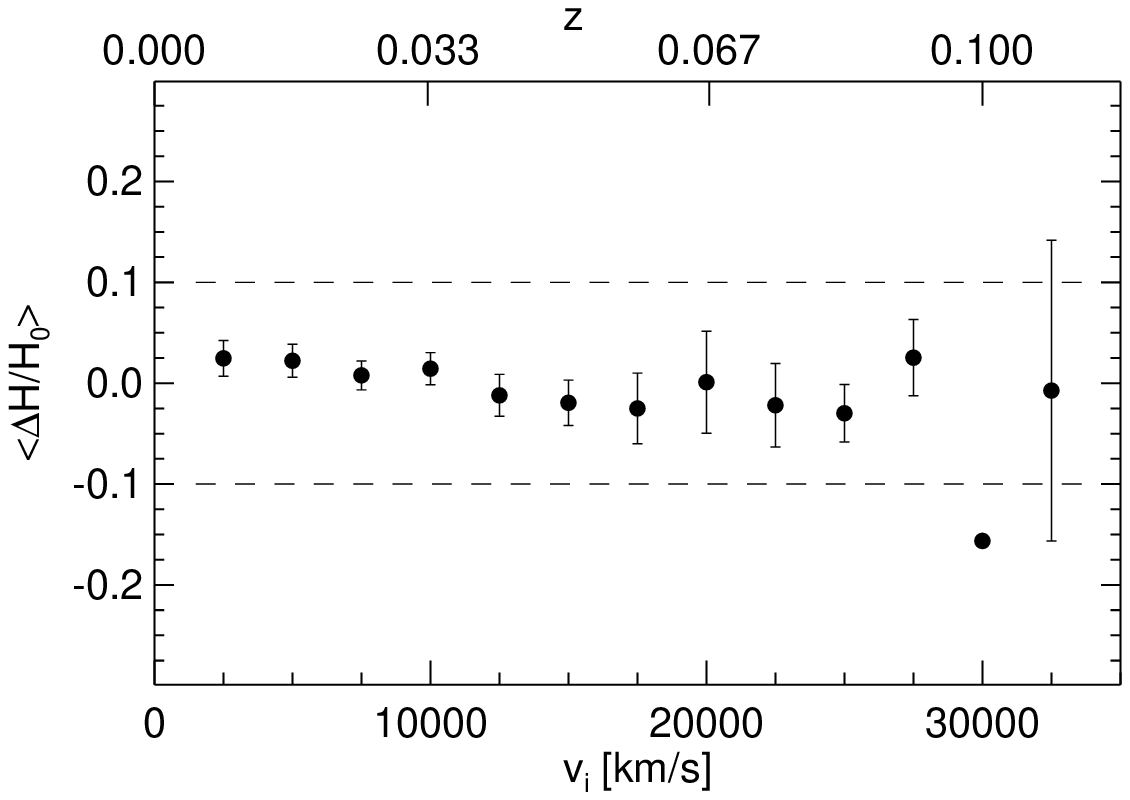,width=\floatwidth}}
  \centerline{\psfig{file=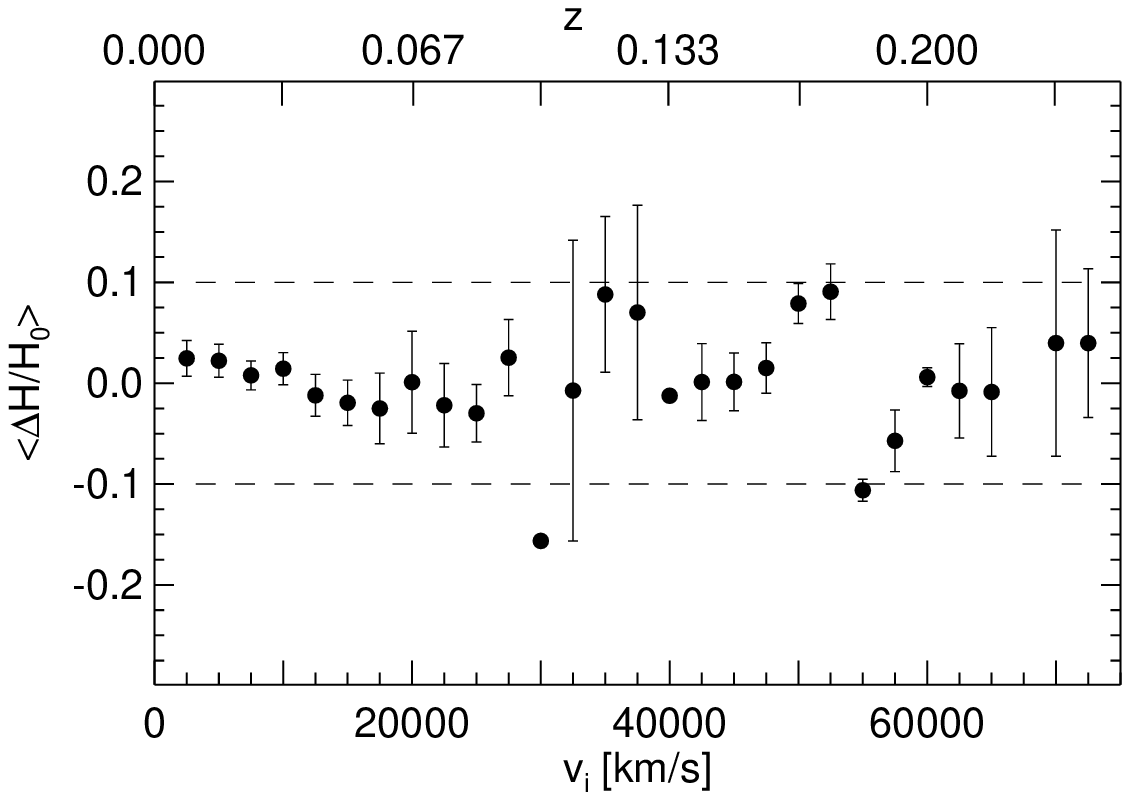,width=\floatwidth}}
  \caption{The variation of $H_0$ with redshift derived from
    relative distances. a) Out to $35\,000\kms$ using relative
    cluster distances, SNe\,Ia, and first-ranked E galaxies in groups
    and rich clusters.
    b) Out to $72\,500\kms$; beyond $35\,000\kms$ the data depend
    only on first-ranked E galaxies in rich clusters.}
  \label{fig:6}
\end{figure}

Inspection of Fig.\,\ref{fig:6} suggests that the
Hubble constant decreases from $1000\kms$ to about $18\,000\kms$
by $\sim5\%$. This trend is independently supported by the
first-ranked cluster galaxies of \citeauthor{Lauer:Postman:94} (1994),
which have not been used here. Beyond $18\,000\kms$ the scatter
becomes large, leaving the possibility of local $\pm10\%$
variations of $H_0$, but the distant overall {\em mean\/} of $H_0$ lies
close to the value found at $\sim10\,000\kms$.

   A proviso should be added. In Figure~\ref{fig:5} different objects
were combined into a single Hubble diagram necessitating a shift for
different absolute magnitudes and introducing additional degrees of
freedom. Correspondingly the behavior of $\Delta H/H_0$ in
Fig.~\ref{fig:6} is not as robust as one would wish. In particular the
{\em position\/} of the minimum at $18\,000\kms$ depends only on a few
points. A well occupied Hubble diagram with only one kind of objects
would provide a stronger test. A continuation of the Cerro Tololo
programme \cite{Hamuy:etal:96} to discover and measure SNe\,Ia at {\em
all distances\/} would therefore be of utmost importance. This will not
only provide  -- if carried to very large distances -- $q_0$ and
$\Lambda$ (cf. \citeauthor{Perlmutter:etal:98} 1998;
\citeauthor{Schmidt:etal:98} 1998), but also to a mapping of the
Hubble flow in function of distance.

   The consequences of deviations from linear expansion on the
determination of $H_0$ are clear.
If our peculiar motion of $630\kms$ is taken to be typical, a
$10\%$ error is to be expected from a single object as far out as
$6000\kms$. In addition $H_0$ determinations within $10\,000\kms$ may
be systematically too large by a few percent and in the range $10\,000
- 20\,000\kms$ too low by the same amount. At $10\,000\kms$ $H_0$ is
apparently closely the same as the mean value of $H_0\;$(cosmic) over
very large scales.

\section{\boldmath$H_0$ through Cepheids and SNe\,Ia}
\label{sec:3}
The most direct and reliable route to $H_0$ became accessible when it
was reliazed that SNe\,Ia at maximum light are very useful standard candles
(\citeauthor{Kowal:68} 1968; Sandage \& Tammann
\citeyear{Sandage:Tammann:82}, \citeyear{Sandage:Tammann:93};
Cadonau, Sandage, \& Tammann \citeyear{Cadonau:etal:85};
\citeauthor{Branch:Tammann:92} 1992; Tammann \& Sandage
\citeyear{Tammann:Sandage:95}; \citeauthor{Hamuy:etal:95}
\citeyear{Hamuy:etal:95}, \citeyear{Hamuy:etal:96}).
If one excludes the few red SNe\,Ia, which are underluminous because
of absorption and/or intrinsic peculiarities, and restricts oneself to
blue, spectroscopically uniform (``Branch-normal'';
Branch, Fisher, \& Nugent \citeyear{Branch:etal:93}) SNe\,Ia with
$(B_{\max} -V_{\max})\le
0.2$, their luminosity scatter is $\sigma_{\rm M}\le 0.25$.
The corresponding Hubble diagram, averaged
over $B_{\max}$ and $V_{\max}$ magnitudes (cf. Fig.~\ref{fig:3}) is
defined by 21 SNe\,Ia which have occurred in spiral galaxies (see
below) and with good photometry after 1985 to be
\begin{equation}\label{equ:1}
   \log cz = 0.2\,m_{\rm B,V} + (0.659\pm0.031),
\end{equation}
which is easily transformed into
\begin{equation}\label{equ:2}
   \log H_0 = 0.2\,M_{\rm B,V} + (5.659\pm0.031),
\end{equation}

   To obtain $H_0$ out to $30\,000\kms$ it is necessary to calibrate
the absolute magnitude of blue SNe\,Ia at maximum light. This is possible
since some SNe\,Ia have been observed in galaxies sufficiently close
to determine their distances by means of their Cepheids. However, this
is possible only since the advent of {\sl HST}.

   For the luminosity calibration of SNe\,Ia a small {\sl HST\/} team
has been formed comprising A.~Sandage, A.~Saha, L.~Labhardt,
F.\,D.~Macchetto, N.~Panagia, \& G.\,A.~Tammann. They have observed so
far the Cepheids in six galaxies having produced seven SNe\,Ia. The
resulting Cepheid distances yield the corresponding values of $M_{\rm
  B,V}$. The latter turn out to be quite {\em uniform\/} confirming
the basic assumption of standard candles. If the resulting mean value
of $M_{\rm B,V} = -19.52\pm0.04$ \cite{Saha:etal:97} for the seven
SNe\,Ia is inserted into equation~(\ref{equ:2}) one obtains
\begin{equation}\label{equ:3}
   H_0\,(30\,000\kms) = 57 \pm 3.
\end{equation}
It must be stressed that the empirical luminosity calibration agrees
within $0\mag1$ with present theoretical models
(cf. \citeauthor{Branch:98} 1998)!

\begin{figure}[t]
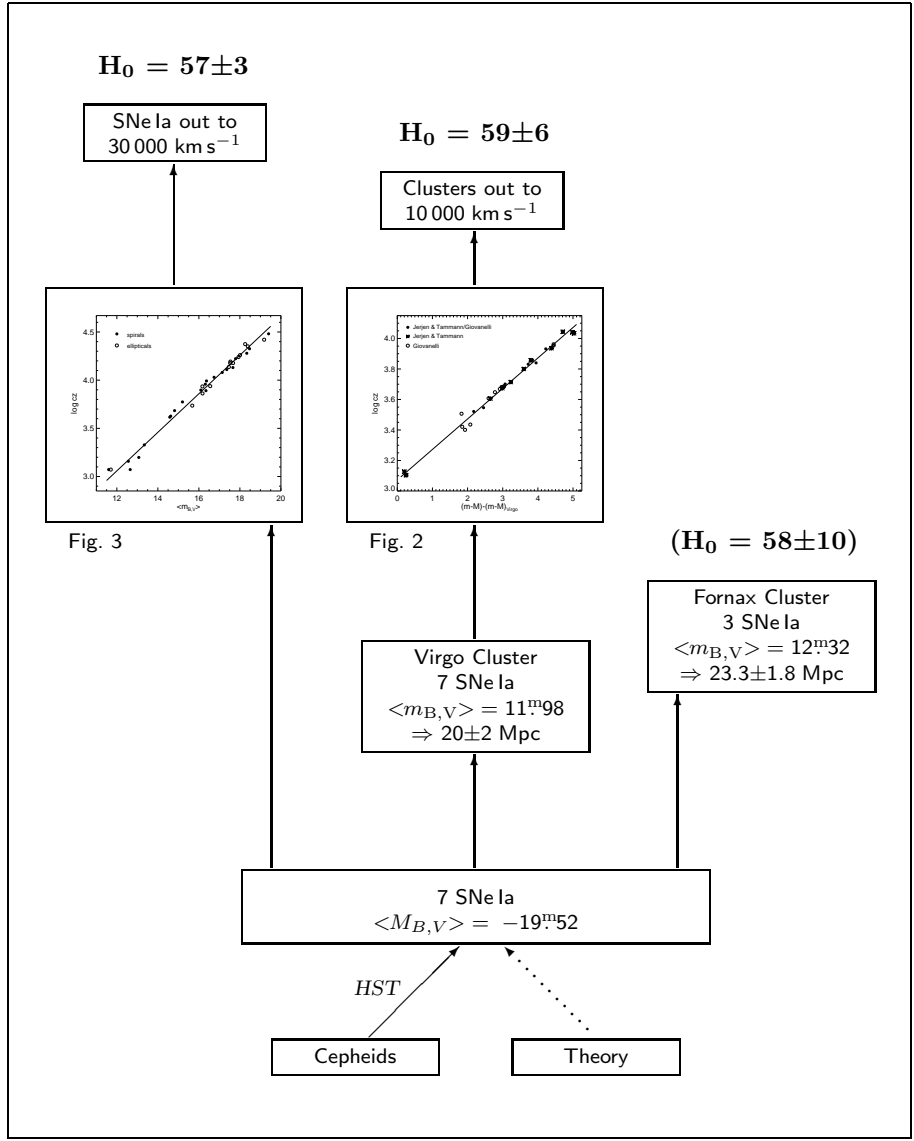

\begin{center}
{\footnotesize \sf
\def\floatwidth{0.25\textwidth}
\setlength{\unitlength}{1mm}
\def\xx{120}
\def\yy{151}
\begin{picture}(\xx,\yy)(0,0)
 \put(0,0){\line(\xx,0){\xx}}
 \put(0,0){\line(0,\yy){\yy}}
 \put(0,\yy){\line(\xx,0){\xx}}
 \put(\xx,0){\line(0,\yy){\yy}}
 \put(22,144){\makebox(0,0)[t]{\normalsize \bf H\boldmath$_0$ $=$ 57$\pm$3}}
 \put(62,135){\makebox(0,0)[t]{\normalsize \bf H\boldmath$_0$ $=$ 59$\pm$6}}
 \put(10,133){\fbox{\parbox{2.2cm}{\centering SNe\,Ia out to \\
  30\,000\kms}}}
 \put(49.5,124){\fbox{\parbox{2.2cm}{\centering Clusters out to \\
 10\,000\kms}}}
 \put(5,83){\fbox{\psfig{file=hubble3.eps,width=\floatwidth}}}
 \put(45,83){\fbox{\psfig{file=hubble2.eps,width=\floatwidth}}}
 \put(22,113.5){\vector(0,0){16}}
 \put(62,113.5){\vector(0,0){7}}
 \put(8,78.5){Fig.~3}
 \put(48,78.5){Fig.~2}
 \put(47,58){\fbox{\parbox{2.8cm}{\centering
   Virgo Cluster \\ 7 SNe\,Ia \\ $<\!\!{m}_{\rm B,V}\!\!>\;=\;$11$\mag$98 \\
   $\Rightarrow$ 20$\pm$2$\;$Mpc}}}
 \put(62,66.5){\vector(0,0){15}}
 \put(85,66){\fbox{\parbox{2.8cm}{\centering
   Fornax Cluster \\ 3 SNe\,Ia \\ $<\!\!{m}_{\rm B,V}\!\!>\;=\;$12$\mag$32 \\
   $\Rightarrow$ 23.3$\pm$1.8$\;$Mpc}}}
 \put(100.5,81){\makebox(0,0)[t]{\normalsize \bf (H\boldmath$_0$ $=$
   58$\pm$10)}}
 \put(31,30){\fbox{\parbox{6cm}{\hspace*{1cm}
\\ \centering 7 SNe\,Ia \\ $<\!\!{M}_{B,V}\!\!>\;=\;-$19$\mag$52}}}
 \put(35,36){\vector(0,0){45.5}}
 \put(62,36){\vector(0,0){15}}
 \put(89,36){\vector(0,0){23}}
 \put(35,10){\fbox{\parbox{2cm}{\centerline{Cepheids}}}}
 \put(67,10){\fbox{\parbox{2cm}{\centerline{Theory}}}}
 \put(46,19){{\sl HST}}
 \put(48,13.5){\vector(1,1){12}}
 \put(77,14.5){\circle*{0.5}}
 \put(76,15.5){\circle*{0.5}}
 \put(75,16.5){\circle*{0.5}}
 \put(74,17.5){\circle*{0.5}}
 \put(73,18.5){\circle*{0.5}}
 \put(72,19.5){\circle*{0.5}}
 \put(71,20.5){\circle*{0.5}}
 \put(70,21.5){\circle*{0.5}}
 \put(69,22.5){\circle*{0.5}}
 \put(68,23.5){\circle*{0.5}}
 \put(68,23.5){\vector(-1,1){2}}
\end{picture}
}
 \end{center}
 \caption{The distance scale built only on SNe\,Ia.}
 \label{fig:7}
\end{figure}

   The above value of $M_{\rm B,V}$ can also be applied to the seven
blue SNe\,Ia in the Virgo cluster and the three SNe\,Ia in the Fornax
cluster (cf. Fig.~\ref{fig:7}). The resulting distance of the Fornax
cluster is not very useful because the peculiar velocity of this
cluster is unknown. If one assumes $v_{\rm Fornax}=1350\pm250\kms$ one
can state only $H_{0}=58\pm10$. The resulting distance of the Virgo
cluster is more significant because this cluster is tightly tied
through relative cluster distances into the Hubble field out to
$10\,000\kms$ (Figure~\ref{fig:2}). The Hubble line in
Figure~\ref{fig:2} implies
\begin{equation}\label{equ:4}
   H_0 = 0.2\,(m-M)_{\rm Virgo} - (8.070\pm0.011).
\end{equation}
If the Virgo modulus from SNe\,Ia is then inserted into
equation~(\ref{equ:4}) one obtains
\begin{equation}\label{equ:5}
   H_0\,(10\,000\kms) = 59 \pm 6
\end{equation}
in perfect agreement with equation~(\ref{equ:3}).

   There has been considerable discussion whether the peak luminosity
of blue SNe\,Ia (slightly) depends on second parameters
(cf. \citeauthor{Saha:etal:97} 1997). There is indeed some dependence
on SN decline rate, SN color, and Hubble type of the parent
galaxy. Almost all of these dependences can be accounted for by allowing
for a luminosity dependence on Hubble type, SNe\,Ia being fainter by
$0\mag18$ in $B$ and $V$ in E/S0 galaxies than their brethren in
spiral galaxies. This effect has been allowed for in the
foregoing. Conservatively the seven calibrating SNe\,Ia have been
assumed to comply with spiral populations (actually two calibrators
are in the Am galaxy NGC\,5253 and are, if anything,
underluminous). Equations (\ref{equ:1}) and (\ref{equ:2}) represent
only the SNe\,Ia in spiral galaxies, and the underluminosity of the
SNe\,Ia in the Fornax cluster, which have occurred in E/S0 galaxies,
has been accounted for.

   Special choices of the second-parameter corrections can drive $H_0$
up to $62$. However, the increase of $H_0$ through second-parameter
corrections is paradoxical because it implies that uncorrected
calibrators are observed too bright and distant SNe\,Ia too faint.
In reality the distant SNe\,Ia must be strongly biased in favor of the
most luminous objects, which are more easily detected and stay longer
above the detection limit. This fundamental effect of stellar
statistics makes the value of $H_0$ in equation~(\ref{equ:3}) an upper
limit (cf. \citeauthor{Tammann:etal:97} 1997).

   The value of $H_0$ in equation~(\ref{equ:3}) depends, of course, on
the goodness of the Cepheid distances. These, however, seem now secure
and uncontroversial with {\em systematic\/} errors of $\le5\%$.
For the {\sl HST\/} observations in $I$ and $V$ the period-luminosity
relations of \cite*{Madore:Freedman:91} has been used.
The zeropoint of the relation is
independently confirmed by the calibration through Galactic clusters
\cite{Sandage:Tammann:71,Feast:95}, the LMC distance derived by other
means (cf. Federspiel, Tammann and Sandage
\citeyear{Federspiel:etal:98}), stellar radii
\cite{DiBenedetto:97}, the Baade-Becker-Wesselink method
\cite{Laney:Stobie:92}, and trigonometric parallaxes
\cite{Madore:Freedman:98,Sandage:Tammann:98}.
--- Attempts to improve
Cepheid distances with the help of a period-luminosity-color (PLC)
relation (e.\,g. \citeauthor{Kochanek:97} 1997) are doomed because
stellar evolution models combined with a pulsation code show that the
basic assumptions going into a PLC relation are not met
\cite{Saio:Gautschy:98}. The same models show that the much-discussed
metallicity has minimal effect on the P-L$_{\rm bol}$ relation;
remaining metallicity effects enter only through the bolometric
correction and the possibly conic shape of the instability strip
\cite{Sandage:98a}.

   Typical errors of individual Cepheid distances derived from $HST$
data are $\pm10\%$ due to the width of the instability strip and
restricted sample size and due to absorption. For five of the seven
calibrating SNe\,Ia the error is smaller because they suffer closely
the same absorption as the Cepheids, and hence only apparent distance
moduli are needed.

\section{\boldmath$H_0$ from the Virgo Cluster and other Methods}
\label{sec:4}
It is in addition possible to build the extragalactic distance scale
independent of the blue SNe\,Ia. This not only provides a valuable
consistency check but also yields a calibration of $H_0$ in its own right.

   The only interdependence of the two routes is the use of Cepheid
distances, but here they do not yield the only basis. The
period-luminosity relation of RR Lyrae stars \cite{Sandage:93} now
confirmed through trigonometric parallaxes from the Hipparcos
satellite \cite{Gratton:etal:97,Reid:98}, and the physical distance
determination of LMC through the ring of SN\,1984A \cite{Panagia:98}
are also involved.

\begin{figure}[t]
\begin{center}
{\footnotesize \sf
\def\floatwidth{0.25\textwidth}
\setlength{\unitlength}{1mm}
\def\xx{120}
\def\yy{151}
\begin{picture}(\xx,\yy)(0,0)
 \put(0,0){\line(\xx,0){\xx}}
 \put(0,0){\line(0,\yy){\yy}}
 \put(0,\yy){\line(\xx,0){\xx}}
 \put(\xx,0){\line(0,\yy){\yy}}
 \put(110,146){\makebox(0,0)[t]{\normalsize \bf
    H\boldmath$_0$ $\approx$ 58}}
 \put(53,146){\makebox(0,0)[t]{\normalsize \bf H\boldmath$_0$ $=$ 55$\pm$4}}
 \put(53,138.5){\vector(0,0){3.5}}
 \put(40.5,134){\fbox{\parbox{2.2cm}{\centering
    Clusters out to \\ 10\,000\kms}}}
 \put(36,91){\fbox{\psfig{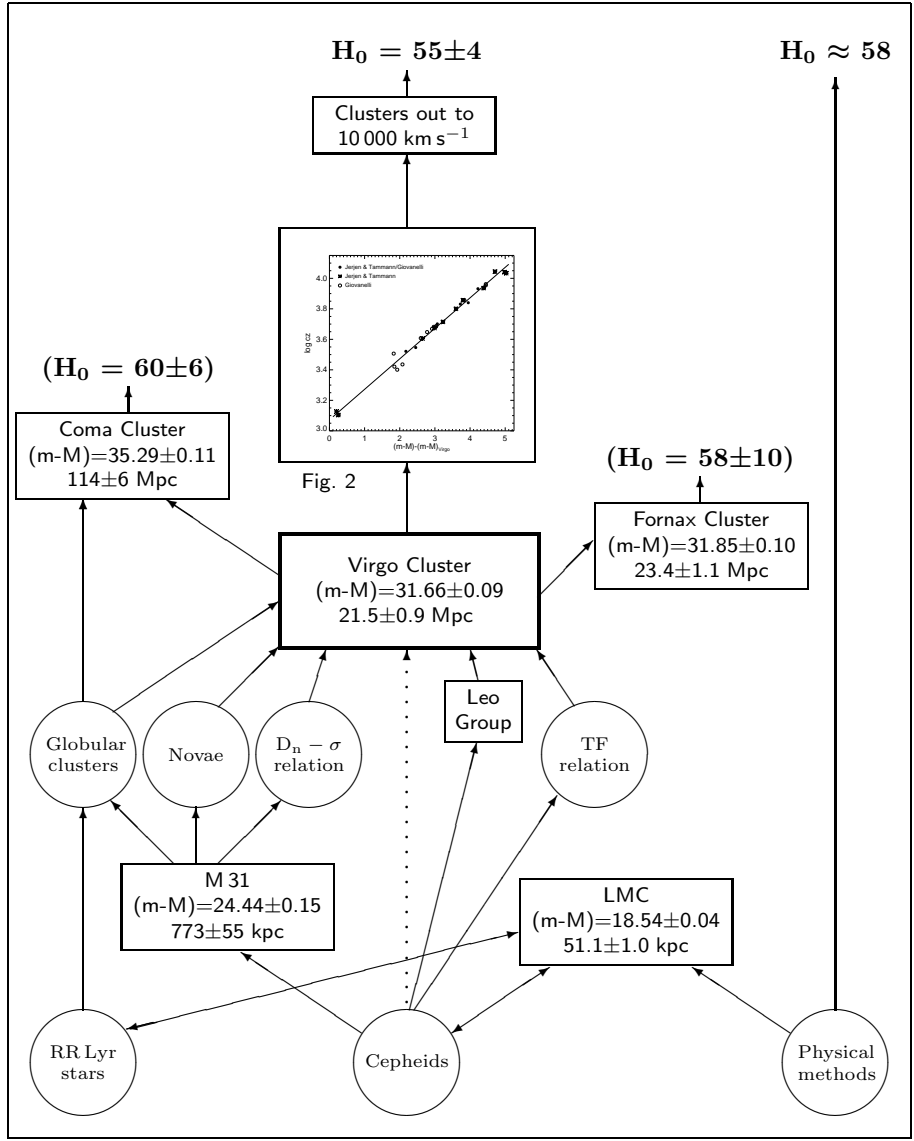}}}
 \put(39,86.5){Fig.~2}
 \put(53,121){\vector(0,0){10}}
 \put(53,80.3){\vector(0,0){9.5}}
 \put(36,74.9){\vector(-3,2){15}}
 \put(70.4,72){\vector(1,1){7.4}}
 \put(16,104){\makebox(0,0)[t]{\normalsize \bf
    (H\boldmath$_0$ $=$ 60$\pm$6)}}
 \put(16,96.5){\vector(0,0){3.5}}
 \put(1,90){\fbox{\parbox{2.6cm}{\centering
    Coma Cluster \\ (m-M)$=$35.29$\pm$0.11 \\
     114$\pm$6$\;$Mpc}}}
 \put(36,72){\fboxrule=1pt \fbox{\parbox{3.2cm}{\centering \vspace*{5pt}
    Virgo Cluster \\  (m-M)$=$31.66$\pm$0.09 \\
    21.5$\pm$0.9$\;$Mpc \vspace*{5pt}}}}
 \put(92,92){\makebox(0,0)[t]{\normalsize \bf (H\boldmath$_0$ $=$
   58$\pm$10)}}
 \put(92,84.5){\vector(0,0){3.5}}
 \put(78,78){\fbox{\parbox{2.6cm}{\centering
   Fornax Cluster \\  (m-M)$=$31.85$\pm$0.10 \\
   23.4$\pm$1.1$\;$Mpc}}}
 \put(58,56){\fbox{\parbox{0.8cm}{\centering
   Leo \\ Group}}}
 \put(62.5,61){\vector(-1,4){1}}
%
\put(14,56.7){\vector(3,2){22}}
\put(28,57.4){\vector(1,1){8}}
\put(40,58){\vector(1,3){2.3}}
\put(75.3,57.5){\vector(-2,3){5}}
\def\py{51}
 \put(10,58){\vector(0,0){27}}
 \put(10,58){\vector(0,0){27}}
\put(10,\py){\circle{40}}
\put(10,\py){\makebox(0,0)[c]{\parbox{1cm}{\centering \scriptsize
      Globular \\ clusters}}}
\put(25,\py){\circle{40}}
\put(25,\py){\makebox(0,0)[c]{\parbox{1cm}{\centering \scriptsize
      Novae}}}
\put(40,\py){\circle{40}}
\put(40,\py){\makebox(0,0)[c]{\parbox{1cm}{\centering \scriptsize
      D$_{\rm n}-\sigma$ \\ relation}}}
\put(25,36.5){\vector(0,0){7.5}}
\put(22,36.5){\vector(-1,1){8.3}}
\put(28,36.5){\vector(1,1){8.3}}
\put(78,\py){\circle{40}}
\put(78,\py){\makebox(0,0)[c]{\parbox{1cm}{\centering \scriptsize
      TF \\ relation}}}
\def\py{30}
 \put(15,\py){\fbox{\parbox[c]{2.6cm}{\centering
   M\,31 \\ (m-M)$=$24.44$\pm$0.15 \\
     773$\pm$55$\;$kpc}}}
 \put(68,28){\fbox{\parbox{2.6cm}{\centering
   LMC \\ (m-M)$=$18.54$\pm$0.04 \\
     51.1$\pm$1.0$\;$kpc}}}
%
\def\py{10}
 \put(10,17){\vector(0,0){27}}
\put(15.8,14.4){\vector(4,1){52}}
\put(19.8,15.4){\vector(-4,-1){4}}
\put(10,\py){\circle{60}}
\put(10,\py){\makebox(0,0)[c]{\parbox{1cm}{\centering \scriptsize
      RR\,Lyr \\ stars}}}
%
\def\dx{53}
 \put(\dx,64){\vector(0,0){1}}
 \put(\dx,62){\circle*{0.5}}
 \put(\dx,60){\circle*{0.5}}
 \put(\dx,58){\circle*{0.5}}
 \put(\dx,56){\circle*{0.5}}
 \put(\dx,54){\circle*{0.5}}
 \put(\dx,52){\circle*{0.5}}
 \put(\dx,50){\circle*{0.5}}
 \put(\dx,48){\circle*{0.5}}
 \put(\dx,46){\circle*{0.5}}
 \put(\dx,44){\circle*{0.5}}
 \put(\dx,42){\circle*{0.5}}
 \put(\dx,40){\circle*{0.5}}
 \put(\dx,38){\circle*{0.5}}
 \put(\dx,36){\circle*{0.5}}
 \put(\dx,34){\circle*{0.5}}
 \put(\dx,32){\circle*{0.5}}
 \put(\dx,30){\circle*{0.5}}
 \put(\dx,28){\circle*{0.5}}
 \put(\dx,26){\circle*{0.5}}
 \put(\dx,24){\circle*{0.5}}
 \put(\dx,22){\circle*{0.5}}
 \put(\dx,20){\circle*{0.5}}
 \put(\dx,18){\circle*{0.5}}
\put(47,14){\vector(-3,2){16}}
\put(59,14){\vector(3,2){13}}
  \put(60.5,15){\vector(-3,-2){1.5}}
\put(54,17){\vector(2,3){19}}
\put(53.4,17){\vector(1,4){8.9}}

\put(\dx,\py){\circle{60}}
\put(\dx,\py){\makebox(0,0)[c]{\scriptsize Cepheids}}
 \put(110,17){\vector(0,0){124}}
 \put(104,14){\vector(-3,2){13}}
\put(110,\py){\circle{60}}
\put(110,\py){\makebox(0,0)[c]{\parbox{1cm}{\centering \scriptsize
      Physical \\ methods}}}
%
\end{picture}
}
  \end{center}
  \caption{The distance scale built on distance indicators other than
           SNe\,Ia. (All errors are internal errors)}
  \label{fig:8}
\end{figure}

   The intertwined network of the second route is schematically shown
in Fig.~\ref{fig:8}. The center piece is the distance of the Virgo
cluster which follows from various methods (Table~\ref{tab:1}).
%
\begin{table}[b]
\caption{The Virgo cluster modulus from various methods}
\label{tab:1}
\begin{center}
\begin{tabular}{lllc}
\noalign{\smallskip}
\hline
\noalign{\smallskip}
        Method               & $(m-M)_{\rm Virgo}$ & Hubble type & Source\\
\noalign{\smallskip}
\hline
\noalign{\smallskip}
        Cepheids              &  $31.52 \pm  0.21$ &     S       & 1 \\
        Tully-Fisher          &  $31.58 \pm  0.24$ &     S       & 2 \\
        Globular Clusters     &  $31.67 \pm  0.15$ &     E       & 3 \\
        D$_{\rm n} - \sigma$  &  $31.85 \pm  0.19$ &     S0, S   & 4 \\
        Novae                 &  $31.46 \pm  0.40$ &     E       & 5 \\
\noalign{\smallskip}
\hline
\noalign{\smallskip}
        Mean:                 &  $31.66 \pm  0.09$ &
 ($\Rightarrow 21.5\pm0.9\;$Mpc) \\
\noalign{\smallskip}
\hline
\noalign{\smallskip}
\end{tabular}
\begin{minipage}{0.9\textwidth}\footnotesize
Sources:
\begin{enumerate}
\def\parsep{0pt}
\def\itemsep{0pt}
\item See text
\item Federspiel, Tammann, \& Sandage 1998\nocite{Federspiel:etal:98}
\item The luminosity function of globular clusters (GC) peaks at
  $M_{\rm B}=-6.90\pm0.11$ and $M_{\rm V}=-7.60\pm0.11$ as determined
  from Galactic GCs with modern RR\,Lyrae star distances {\em and\/}
  Cepheid-calibrated GCs in M\,31 \cite{Sandage:Tammann:95}. Apparent
  peak magnitudes of Virgo members from various authors are compiled
  by \citeauthor{Sandage:Tammann:95} (1995; cf. also
  \citeauthor{Tammann:Federspiel:97} 1997).
\item A reasonably tight D$_{\rm n}-\sigma$ relation of S0 and spiral
  galaxies in the Virgo cluster has been published by
  \cite*{Dressler:87}. The zeropoint calibration rests on the distance
  of the Galactic bulge ($7.8\;$kpc) and the Cepheid distances of
  M\,31 and M\,81 \cite{Sandage:Tammann:88,Tammann:88}.
\item \cite*{Pritchet:vandenBergh:87} found from six novae in Virgo
  cluster ellipticals that that they are $7\mag0\pm0\mag4$ more
  distant than the {\em apparent\/} distance modulus of M\,31 of
  $(m-M)_{\rm AB}=24.58\pm0.10$ from Cepheids
  \cite{Madore:Freedman:91} {\em and\/} Galactic novae
  \cite{Capaccioli:etal:89}. \cite*{Livio:97} found from a
  semi-theoretical analysis of the six Virgo novae $(m-M)_{\em
  Virgo}=31.35\pm0.35$. A low-weight mean of $31.46$ is adopted.
\end{enumerate}
\end{minipage}
\end{center}
\end{table}

A special note is required on the Virgo cluster distance from
Cepheids. There are now three bona fide cluster members 
and two outlying members with Cepheid
distances from {\sl HST\/} (Table~\ref{tab:2};
cf. \citeauthor{Freedman:etal:98} 1998).
%
\begin{table}[t]
\caption{The Virgo cluster members with Cepheid distances}
\label{tab:2}
\begin{center}
\begin{minipage}{0.87\textwidth}
\begin{tabular}{llll}
\noalign{\smallskip}
\hline
\noalign{\smallskip}
  Galaxy & $(m-M)_{\rm Cepheids}$ & Remarks & $(m-M)_{\rm TF}$ \\
\noalign{\smallskip}
\hline
\noalign{\smallskip}
  NGC\,4321$^*$ & $31.04\pm0.21$ & highly resolved & $31.21\pm0.40$ \\
  NGC\,4496A$^{*\,*}$
            & $31.13\pm0.10$ & highly resolved & $30.67\pm0.40$ \\
  NGC\,4536$^{*\,*}$
            & $31.10\pm0.13$ & highly resolved & $30.72\pm0.40$ \\
  NGC\,4571 & $30.87\pm0.15$ & extremely resolved & $31.75\pm0.40$ \\
  NGC\,4639 & $32.03\pm0.23$ & poorly resolved & $32.53\pm0.40$ \\
\noalign{\smallskip}
\hline
\noalign{\smallskip}
\end{tabular}
\mpspacing{0.8}
{\footnotesize
\hspace*{5pt}$^*$~From a re-analysis of the {\sl HST\/} observations
 \cite*{Narasimha:Mazumdar:98} obtained $(m-M)=31.55\pm0.28$. \\
$^{*\,*}$~In the W-cloud outside the confidence
 boundaries of the Virgo cluster (cf. Federspiel \etal 1998).}
\end{minipage}
\end{center}
\end{table}
The wide range of their distance moduli, corresponding to $14.9$ to
$25.5\;$Mpc, reveals the important depth effect of the cluster. The
first four galaxies in Table~\ref{tab:2} have been chosen from the
atlas of \cite*{Sandage:Bedke:88} because they are highly resolved and
seemed easy as to their Cepheids. They are therefore {\em expected\/}
to lie on the near side of the cluster. In contrast NGC\,4639 has been
chosen as parent to SN\,1990N and hence independently of its distance;
correspondingly this distance is expected to be statistically more
representative.
A straight mean of the distances in Table~\ref{tab:2} is
therefore likely to be an underestimate. Indeed the mean Tully-Fisher
(TF) distance modulus of the five galaxies is $0\mag2$ (corresponding to
$10\%$ in distance) {\em smaller\/} than the mean distance of a
complete and fair sample of TF distances \cite{Federspiel:etal:98}.

   A preliminary Cepheid distance of the Virgo cluster is obtained by
taking the Cepheid distance of the Leo group of $(m-M)=30.27\pm0.12$,
based now on three galaxies with Cepheids from {\sl HST}, and to step
up this value by the modulus difference of $\Delta(m-M)=1.25\pm0.13$
\cite{Tammann:Federspiel:97} between the Leo group and the Virgo
cluster. The corresponding result is shown in Table~\ref{tab:1}.

   If the adopted Virgo cluster modulus in Table~\ref{tab:1} is
inserted into equation~(\ref{equ:1}) one obtains
\begin{equation}\label{equ:6}
   H_0\,(10\,000\kms) = 55 \pm 4.
\end{equation}

   The Virgo cluster distance modulus can be used to derive the
distances of the Fornax and Coma clusters by adding the respective
modulus differences to the former. The relevant modulus differences
are compiled in Tables~\ref{tab:3} and \ref{tab:4}. Although there are
strong reservations against distances from planetary nebulae and
surface brightness fluctuations the {\em differential\/} distances
from these methods have been included here, because they are
independent of the zeropoint calibration and less sensitive to
selection effects. Also the absolute Coma cluster distance from the
surface brightness fluctuation method has been tentatively
included. The globular cluster distance of the Coma cluster is based
on the same zeropoint calibration as the Virgo cluster
(cf. Table~\ref{tab:1}).
%
\begin{table}[t]
\caption{The relative distance Fornax - Virgo cluster (E/S0 galaxies)}
\label{tab:3}
\begin{center}
\begin{tabular}{rll}
\noalign{\smallskip}
\hline
\noalign{\smallskip}
  \multicolumn{1}{c}{$\Delta(m-M)_{{\rm For}-{\rm Vir}}$} & Method  &
           Reference \\
\noalign{\smallskip}
\hline
\noalign{\smallskip}
  $0.44\pm0.30$ & first-ranked galaxy      &
    \citeauthor{Sandage:etal:76} 1976 \\
  $0.14\pm0.16$ & D$_{n}-\sigma$ method    &
    \citeauthor{Faber:etal:89} 1989 \\
  $0.28\pm0.08$ & surface brightness &
    \citeauthor{Jerjen:Binggeli:97} 1997 \\
  $0.20\pm0.08$ & surf. brightness fluctuations &
    \citeauthor{Tonry:97} 1997 \\
  $0.32\pm0.10$ & planetary nebulae & \citeauthor{Jacoby:97} 1997 \\
 $-0.09\pm0.12$ & globular clusters & \citeauthor{Whitmore:97} 1997 \\
\noalign{\smallskip}
\hline
\noalign{\smallskip}
  $0.19\pm0.04$ & mean \hspace*{67pt} $\Rightarrow$ &
   $(m-M)_{\rm For}=31.85\pm0.10$   \\
\noalign{\smallskip}
\hline
\end{tabular}
\end{center}
\end{table}
%
\begin{table}[t]
\caption{The distance of the Coma cluster (E/S0 galaxies)}
\label{tab:4}
\begin{center}
\begin{tabular}{rll}
\noalign{\smallskip}
\hline
\noalign{\smallskip}
    $(m-M)_{\rm Coma}$ & Method  &  Reference \\
\noalign{\smallskip}
\hline
\noalign{\smallskip}
  $>35.22\pm0.20$ & globular clusters     &
    \citeauthor{Baum:etal:95} 1995\\
   $35.34\pm0.20$ & globular clusters & \citeauthor{Baum:etal:97} 1997 \\
   $35.04\pm0.31$ & surf. brightness fluctuations &
    \citeauthor{Thomsen:etal:97} 1997 \\
\noalign{\smallskip}
   \multicolumn{2}{l}{$\Delta (m-M)_{{\rm Coma}-{\rm Vir}}$} & \\
\noalign{\smallskip}
  $3.34\pm0.30$ & first-ranked galaxy &
    \citeauthor{Sandage:etal:76} 1976 \\
  $4.16\pm0.20$ & 10 brightest galaxies & \citeauthor{Weedman:76} 1976 \\
  $3.74\pm0.14$ & D$_{n}-\sigma$ method    &
    \citeauthor{Faber:etal:89} 1989 \\
\noalign{\medskip}
  $3.74\pm0.20$ & mean difference \hspace*{30pt} $\Rightarrow$ &
    $(m-M)_{\rm Coma}=35.40\pm0.22$
    \\
\noalign{\smallskip}
\hline
\noalign{\smallskip}
  $35.29\pm0.11$ & overall mean \hspace*{43pt} $\Rightarrow$ &
   $114\pm6\;$Mpc \\
\noalign{\smallskip}
\hline
\end{tabular}
\end{center}
\end{table}

   The resulting values of $H_0$ from the Fornax and Coma clusters are
shown in Figure~\ref{fig:8}. They carry larger errors than $H_0$ from
the Virgo cluster because the former are poorly tied into the cosmic
expansion field. Cosmic velocities of $v_{\rm Fornax}=1350\pm250\kms$ and
$v_{\rm Coma}=7000\pm650\kms$ have been assumed. The Fornax cluster
offers the additional problem that the spiral ``members'' may be
closer than the E/S0 cluster galaxies \cite{Tammann:Federspiel:97}. It
is, however, satisfactory that the independent distances of the Fornax
cluster in Figures~\ref{fig:7} and 8 are closely the same.

   There are many determinations of $H_0$ from {\em field galaxies\/}
in the literature. The difficulty here is that catalogs of field
galaxies are limited by {\em apparent\/} magnitude, which causes the
mean galaxian luminosities to increase with distance because of the
progressive loss of the less luminous galaxies. The corresponding
Malmquist corrections are difficult to apply; their neglect leads
always to too high values of $H_0$.
   Recent determinations of $H_0$ from bias-corrected field galaxies
are compiled in Table~\ref{tab:4b}.
%
\begin{table}[t]
\caption{$H_{0}$ from bias corrected field galaxies}
\label{tab:4b}
\begin{center}
\begin{tabular}{lll}
\noalign{\smallskip}
\hline
\noalign{\smallskip}
         Method       &   \multicolumn{1}{c}{$H_{0}$}      &    Source \\
\noalign{\smallskip}
\hline
\noalign{\smallskip}
Tully-Fisher                           &   $< 60$    &
   \citeauthor{Sandage:94} 1994   \\
M\,101 look-alike diameters            & $43 \pm 11$ &
   \citeauthor{Sandage:93a} \citeyear{Sandage:93a} \\
M\,31 look-alike diameters             & $45 \pm 12$ &
   \citeauthor{Sandage:93b} \citeyear{Sandage:93b}  \\
Spirals with luminosity classes        & $56 \pm  5$ &
   \citeauthor{Sandage:96a} \citeyear{Sandage:96a}  \\
M\,101, M\,31 look alike luminosities  & $55 \pm  5$ &
   \citeauthor{Sandage:96b} \citeyear{Sandage:96b} \\
Tully-Fisher                           & $55 \pm  5$ &
   \citeauthor{Theureau:etal:97} 1997 \\
Galaxy diameters                       & $50 - 55$   &
   \citeauthor{Goodwin:etal:97} 1997 \\
Tully-Fisher                           & $60 \pm 5$  &
   \citeauthor{Federspiel:98} 1998 \\
\noalign{\smallskip}
\hline
\noalign{\smallskip}
mean                                   & $53\;(\pm 3)$  & \\
\noalign{\smallskip}
\hline
\noalign{\smallskip}
\end{tabular}
\end{center}
\end{table}
%

   Field galaxies offer the advantage of full-sky coverage outside the
zone of avoidance. But they are not only the most difficult route to
$H_0$, but also the least satisfactory, having their main thrust as
close as $1000-3000\kms$, i.\,e.\ at a distance where $H_{0}$\,(local)
may still be a few percent higher than $H_{0}$\,(cosmic)
(cf. Fig.~\ref{fig:6}).

\newpage
\section{Physical Determinations of \boldmath$H_0$}
\label{sec:5}
One distinguishes between
astronomical and physical distance determinations. The former depend
always on some adopted distance of a celestial body, be it only the
Astronomical Unit in the case of trigonometric parallaxes. Physical
methods derive the distance solely from the observed physical or
geometrical properties of a specific object.

   In the foregoing use has been made to physical luminosity and
distance determinations of the SN\,1987A remnant, Cepheids, RR\,Lyr
stars, and SNe\,Ia. But in addition there are a number of physical
distance determinations which lead to the value of $H_0$ over very
large scales. They are still model-dependent, but as the number of
objects increases and the models improve, their weight is steadily
increasing.

   For brevity the most recent physical determinations of $H_0$ are
compiled in Table~\ref{tab:5}.
Following \citeauthor{Rephaeli:Yankovitch:97}~(1997) all previous
values of $H_0$ from the SZ effect should be lowered by $\sim\! 10$
units due to relativistic effects.

%
\begin{table}[t]
\caption{$H_0$ from Physical Methods}
\label{tab:5}
\begin{center}
\begin{minipage}{0.9\textwidth}
\begin{center}
\begin{tabular}{llc}
\noalign{\smallskip}
\hline
\noalign{\smallskip}
       Method          &   \hspace*{1ex}$H_{0}$   &   Source$^*$
                                                             \\
\noalign{\smallskip}
\hline
\noalign{\smallskip}
  Sunyaev-Zeldovich effect                    &             &      \\
         \hspace*{1cm} for cluster A 2218     & $45 \pm 20$ &  (1) \\
         \hspace*{1cm} for 6 other clusters   & $60 \pm 15$ &  (2) \\
         \hspace*{1cm} cluster A 2163  &  $78\,(+54,\,-28)$ &  (3) \\
         \hspace*{1cm} 2 clusters             & $42 \pm 10$ &  (4) \\
         \hspace*{1cm} 3 clusters             & $54 \pm 14$ &  (5) \\
         \hspace*{1cm} incl. relativ. effects & $44 \pm  7$ &  (6) \\
\noalign{\smallskip}
  Gravitational lenses                        &             &      \\
         \hspace*{1cm} QSO 0957 + 561         & $62 \pm 7$  & (7) \\
         \hspace*{1cm} B 0218 + 357           & $52-82$     & (8) \\
         \hspace*{1cm} PG 1115 + 080          & $60 \pm 17$ & (9) \\
         \hspace*{1.9cm} ''                   & $52 \pm 14$ & (10) \\
         \hspace*{1.9cm} ''                   & $62 \pm 20$ & (11) \\
\noalign{\smallskip}
  MWB fluctuation spectrum                    & $58 \pm 11$ & (12) \\
         \hspace*{1.9cm} ''                   & $47 \pm  6$ & (13) \\
\noalign{\smallskip}
\hline
\noalign{\smallskip}
\end{tabular}
\end{center}

\mpspacing{0.85}

{\normalsize\footnotesize
        \hspace*{5pt}$^*$\,Sources:
    (1) \citeauthor{McHardy:etal:90} 1990;
        \citeauthor{Birkinshaw:Hughes:94} 1994;
        \citeauthor{Lasenby:Hancock:95} 1995
    (2) \citeauthor{Rephaeli:95} 1995;
        \citeauthor{Herbig:etal:95} 1995
    (3) \citeauthor{Holzapfel:etal:97} 1997
    (4) \citeauthor{Lasenby:Jones:97} 1997
    (5) \citeauthor{Myers:etal:97} 1997
    (6) \citeauthor{Rephaeli:Yankovitch:97} 1997
    (7) \citeauthor{Falco:etal:97} 1997
    (8) \citeauthor{Nair:96} 1996
    (9) \citeauthor{Keeton:Kochanek:97} 1997
   (10) \citeauthor{Kundic:etal:97} 1997
   (11) \citeauthor{Schechter:etal:97} 1997
   (12) \citeauthor{Lineweaver:98} 1998
   (13) \citeauthor{Webster:etal:98} 1998.
}
\end{minipage}\end{center}
\end{table}

   The overall impression from the values in Table~\ref{tab:5} is that
$H_0$ will settle around $H_0\approx58$.

\section{Conclusions}
\label{sec:6}
Two separate routes to the Hubble constant require consistently
$55<H_{0}<60$. The one is based on Cepheid-calibrated blue SNe\,Ia and
their Hubble diagram out to $30\,000\kms$. The other relies on the
Virgo cluster distance from various methods and relative cluster
distances out to $10\,000\kms$. The conclusion that $H_{0}\,({\rm
  cosmic})=57\pm7$ (external error) is in perfect agreement with
present physical distance determinations like the Sunyaev-Zeldovich
effect, gravitational lenses, and the CMB fluctuation spectrum.

   It is not understood in the face of this evidence why values of
$H_0$ as high as $\sim\!70$ are still occasionally proposed
(cf. \citeauthor{Freedman:etal:97} 1997; \citeauthor{Freedman:etal:98}
1998). Principal reasons are the neglect of the Malmquist correction
for selection effects and the adopted too small Virgo cluster
distance. The latter is based
only on the Cepheid distance of M\,100 \cite{Ferrarese:etal:96}, which
actually lies on the {\em near side\/} of the Virgo cluster
(cf. Table~\ref{tab:2}). The small Virgo distance is only seemingly
supported by planetary nebulae \cite{Ciardullo:etal:98} and the
surface brightness fluctuation method \cite{Tonry:97}, i.\,e.\ two
distance indicators which have been included in Table~\ref{tab:3} for
the determination of {\em relative\/} distances, but have utterly
failed for {\rm absolute\/} distances in several cases
(cf. \citeauthor{Tammann:98} 1998).

   The determination of $H_0$ through blue SNe\,Ia has presently the
highest weight. Forthcoming additional calibrators with Cepheid
distances from {\sl HST\/} will further strengthen the calibration. If
{\em more\/} SNe\,Ia will become available with redshifts up to
$\sim30\,000\kms$ and good photometry near maximum, they will lead to
an even tighter Hubble diagram. In this way it will be possible
not only to better map the variations of $H_0$ with distance, but also
to push the error of $H_0$ well below $10\%$.

\mpspacing{0.9}
\bigskip
\noindent {\small {\bf Acknowledgement.}
Support of the Swiss National Science Foundation
is gratefully acknowledged.\ The author thanks the hidden team in
Baltimore who make $HST$ so successfully operational.\
He also thanks Mr.~Bernd Reindl for computational help and the outlay
of the manuscript.}

\mpspacing{1.0}
\clearpage
%
%
%

\begin{thebibliography}{xx}

\bibitem[\protect\citeauthoryear{Baum{\em ~et~al.}}{1997}]{Baum:etal:97}
Baum, W.~A., Hammergren, M., Thomsen, B., Groth, E.~J., Faber, S.~M.,
  Grillmair, C.~J., \& Ajhar, E.~A. 1997, {\em AJ} {\bf 113},~1483.

\bibitem[\protect\citeauthoryear{Baum{\em ~et~al.}}{1995}]{Baum:etal:95}
Baum, W.~A.,{\em ~et~al.} 1995, {\em AJ} {\bf 110},~2537.

\bibitem[\protect\citeauthoryear{Birkinshaw \&
  Hughes}{1994}]{Birkinshaw:Hughes:94}
Birkinshaw, M. \& Hughes, J.~P. 1994, {\em ApJ} {\bf 420},~33.

\bibitem[\protect\citeauthoryear{Branch}{1998}]{Branch:98}
Branch, D. 1998, {\em ARA\&A}, in press.

\bibitem[\protect\citeauthoryear{Branch, Fisher, \&
  Nugent}{1993}]{Branch:etal:93}
Branch, D., Fisher, A., \& Nugent, P. 1993, {\em AJ} {\bf 106},~2383.

\bibitem[\protect\citeauthoryear{Branch \& Tammann}{1992}]{Branch:Tammann:92}
Branch, D. \& Tammann, G.~A. 1992, {\em ARA\&A} {\bf 30},~359.

\bibitem[\protect\citeauthoryear{Cadonau, Sandage, \&
  Tammann}{1985}]{Cadonau:etal:85}
Cadonau, R., Sandage, A., \& Tammann, G.~A. 1985, in: {\em Supernovae as
  Distance Indicators}, ed. N.~Bartel, {\em Lecture Notes in Physics}, {\bf
  224}, Berlin: Springer, p.~15.

\bibitem[\protect\citeauthoryear{Capaccioli{\em
  ~et~al.}}{1989}]{Capaccioli:etal:89}
Capaccioli, M., Della~Valle, M., Rosino, L., \& D'Onofrio, M. 1989, {\em AJ}
  {\bf 97},~1622.

\bibitem[\protect\citeauthoryear{Ciardullo{\em
  ~et~al.}}{1998}]{Ciardullo:etal:98}
Ciardullo, R., Jacoby, G.~H., Feldmeier, J.~J., \& Bartlett, R.~E. 1998, {\em
  ApJ} {\bf 492},~62.

\bibitem[\protect\citeauthoryear{{Di Benedetto}}{1997}]{DiBenedetto:97}
{Di Benedetto}, G.~P. 1997, {\em ApJ} {\bf 486},~60.

\bibitem[\protect\citeauthoryear{Dressler}{1987}]{Dressler:87}
Dressler, A. 1987, {\em ApJ} {\bf 317},~1.

\bibitem[\protect\citeauthoryear{Faber{\em ~et~al.}}{1989}]{Faber:etal:89}
Faber, S., Wegner, G., Burstein, D., Davies, R., Dressler, A., Lynden-Bell, D.,
  \& Terlevich, R. 1989, {\em ApJS} {\bf 69},~763.

\bibitem[\protect\citeauthoryear{Falco, Leh{\'a}r, \&
  Shapiro}{1997}]{Falco:etal:97}
Falco, E.~E., Leh{\'a}r, J., \& Shapiro, I.~I. 1997, {\em AJ} {\bf 113},~540.

\bibitem[\protect\citeauthoryear{Feast}{1995}]{Feast:95}
Feast, M. 1995, {\em Astrophys. Applications of Stellar Pulsations}, {\em ASP
  Conf. Series}, {\bf 83}, p.~209.

\bibitem[\protect\citeauthoryear{Federspiel}{1998}]{Federspiel:98}
Federspiel, M. 1998, in preparation.

\bibitem[\protect\citeauthoryear{Federspiel, Tammann, \&
  Sandage}{1998}]{Federspiel:etal:98}
Federspiel, M., Tammann, G.~A., \& Sandage, A. 1998, {\em ApJ} {\bf 495},~115.

\bibitem[\protect\citeauthoryear{Ferrarese{\em
  ~et~al.}}{1996}]{Ferrarese:etal:96}
Ferrarese, L.,{\em ~et~al.} 1996, {\em ApJ} {\bf 464},~568.

\bibitem[\protect\citeauthoryear{Freedman, Madore, \&
  Kennicutt}{1997}]{Freedman:etal:97}
Freedman, W.~L., Madore, B.~F., \& Kennicutt, C. 1997, in: {\em The
  Extragalactic Distance Scale}, eds. M.~Livio, M.~Donahue, \& N.~Panagia,
  Cambridge: Cambridge University Press, p.~171.

\bibitem[\protect\citeauthoryear{Freedman{\em
  ~et~al.}}{1998}]{Freedman:etal:98}
Freedman, W.~L., Mould, J.~R., Kennicutt, R.~C., \& Madore, B.~F. 1998, in:
  {\em Cosmological Parameters and Evolution of the Universe}, ed. K.~Sato,
  {\em I.\,A.\,U. Symp.}, {\bf 183}, in press.

\bibitem[\protect\citeauthoryear{Giovanelli}{1997}]{Giovanelli:97a}
Giovanelli, R. 1997, in: {\em The Extragalactic Distance Scale}, eds.
  M.~Livio, M.~Donahue, \& N.~Panagia, Cambridge: Cambridge University Press,
  p.~113.

\bibitem[\protect\citeauthoryear{Goodwin{\em ~et~al.}}{1997}]{Goodwin:etal:97}
Goodwin, S.~P., Gribbin, J., \& Hendry, M.~A. 1997, {\em AJ} {\bf 114},~2212.

\bibitem[\protect\citeauthoryear{Gratton{\em ~et~al.}}{1997}]{Gratton:etal:97}
Gratton, R.~G., {Fusi Perci}, F., Carretta, E., Clementini, G., Corsi, C.~F.,
 \& Lattanzi, M. 1997, {\em ApJ} {\bf 491},~749.

\bibitem[\protect\citeauthoryear{Hamuy{\em ~et~al.}}{1995}]{Hamuy:etal:95}
Hamuy, M., Phillips, M.~M., Maza, J., Suntzeff, N.~B., Schommer, R.~A., \&
  Aviles, R. 1995, {\em AJ} {\bf 109},~1.

\bibitem[\protect\citeauthoryear{Hamuy{\em ~et~al.}}{1996}]{Hamuy:etal:96}
Hamuy, M., Phillips, M.~M., Maza, J., Suntzeff, N.~B., Schommer, R.~A., \&
  Aviles, R. 1996, {\em AJ} {\bf 112},~2398.

\bibitem[\protect\citeauthoryear{Herbig, Lawrence, \&
  Readhead}{1995}]{Herbig:etal:95}
Herbig, T., Lawrence, C.~R., \& Readhead, A. C.~S. 1995, {\em ApJ} {\bf
  449},~L5.

\bibitem[\protect\citeauthoryear{Holzapfel{\em
  ~et~al.}}{1997}]{Holzapfel:etal:97}
Holzapfel, W.~L.,{\em ~et~al.} 1997, {\em ApJ} {\bf 480},~449.

\bibitem[\protect\citeauthoryear{Jacoby}{1997}]{Jacoby:97}
Jacoby, G.~H. 1997, in: {\em The Extragalactic Distance Scale}, eds. M.~Livio,
  M.~Donahue, \& N.~Panagia, Cambridge: Cambridge University Press, p.~186.

\bibitem[\protect\citeauthoryear{Jerjen \&
  Binggeli}{1997}]{Jerjen:Binggeli:97}
Jerjen, H. \& Binggeli, B. 1997, {\em The Nature of Elliptical Galaxies},
  {\em ASP Conference Series}, {\bf 116}, p.~298.

\bibitem[\protect\citeauthoryear{Jerjen \& Tammann}{1993}]{Jerjen:Tammann:93}
Jerjen, H. \& Tammann, G.~A. 1993, {\em A\&A} {\bf 276},~1.

\bibitem[\protect\citeauthoryear{Keeton \&
  Kochanek}{1997}]{Keeton:Kochanek:97}
Keeton, C.~R. \& Kochanek, C.~S. 1997, {\em ApJ} {\bf 487},~42.

\bibitem[\protect\citeauthoryear{Kochanek}{1997}]{Kochanek:97}
Kochanek, C.~S. 1997, {\em ApJ} {\bf 491},~13.

\bibitem[\protect\citeauthoryear{Kowal}{1968}]{Kowal:68}
Kowal, C.~T. 1968, {\em AJ} {\bf 73},~1021.

\bibitem[\protect\citeauthoryear{Kundi{\'c}{\em
  ~et~al.}}{1997}]{Kundic:etal:97}
Kundi{\'c}, T., Cohen, J.~G., Blanford, R.~D., \& Lubin, L.~M. 1997, {\em AJ}
  {\bf 114},~507.

\bibitem[\protect\citeauthoryear{Laney \& Stobie}{1992}]{Laney:Stobie:92}
Laney, C.~D. \& Stobie, R.~S. 1992, in: {\em Variable Stars and Galaxies},
  ed. B.~Warner, {\em ASP Conf. Series}, {\bf 30}, p.~119.

\bibitem[\protect\citeauthoryear{Lasenby \&
  Hancock}{1995}]{Lasenby:Hancock:95}
Lasenby, A.~N. \& Hancock, S. 1995, in: {\em Current Topics in
  Astrofundamental Physics: The Early Universe}, eds. N.~S{\'a}nchez \&
  A.~Zichichi, {\em NATO Advanced Science Institutes Series}, {\bf 467},
  Dordrecht: Kluwer Academic Publishers, p.~327.

\bibitem[\protect\citeauthoryear{Lasenby \& Jones}{1997}]{Lasenby:Jones:97}
Lasenby, A.~N. \& Jones, M.~E. 1997, in: {\em The Extragalactic Distance
  Scale}, eds. M.~Livio, M.~Donahue, \& N.~Panagia, Cambridge: Cambridge
  University Press, p.~76.

\bibitem[\protect\citeauthoryear{Lauer \& Postman}{1994}]{Lauer:Postman:94}
Lauer, T.~R. \& Postman, M. 1994, {\em ApJ} {\bf 425},~418.

\bibitem[\protect\citeauthoryear{Lineweaver}{1998}]{Lineweaver:98}
Lineweaver, C.~H. 1998, in: {\em Cosmological Parameters and Evolution of the
  Universe}, ed. K.~Sato, {\em I.\,A.\,U. Symp.}, {\bf 183}, in press.

\bibitem[\protect\citeauthoryear{Livio}{1997}]{Livio:97}
Livio, M. 1997, in: {\em The Extragalactic Distance Scale}, eds. M.~Livio,
  M.~Donahue, \& N.~Panagia, Cambridge: Cambridge University Press, p.~186.

\bibitem[\protect\citeauthoryear{Madore \&
  Freedman}{1991}]{Madore:Freedman:91}
Madore, B. \& Freedman, W.~L. 1991, {\em PASP} {\bf 103},~933.

\bibitem[\protect\citeauthoryear{Madore \&
  Freedman}{1998}]{Madore:Freedman:98}
Madore, B. \& Freedman, W.~L. 1998, {\em ApJ} {\bf 492},~110.

\bibitem[\protect\citeauthoryear{McHardy{\em ~et~al.}}{1990}]{McHardy:etal:90}
McHardy, L.~M., Stewart, G.~C., Edge, A.~C., Cooke, B.~A., Yamashita, K.,\&
  Hatsukade, I. 1990, {\em MNRAS} {\bf 242},~215.

\bibitem[\protect\citeauthoryear{Myers{\em ~et~al.}}{1997}]{Myers:etal:97}
Myers, S.~T., Baker, J.~E., Readhead, A. C.~S., Leitch, E.~M., \& Herbig, T.
  1997, {\em ApJ} {\bf 485},~1.

\bibitem[\protect\citeauthoryear{Nair}{1996}]{Nair:96}
Nair, S. 1996, in: {\em Astrophysical Applications of Gravitational Lensing},
  eds. C.~S. Kochanek \& J.~N. Hewitt, Dordrecht: Kluwer, p.~197.

\bibitem[\protect\citeauthoryear{Narasimha \&
  Mazumdar}{1998}]{Narasimha:Mazumdar:98}
Narasimha, D. \& Mazumdar, A. 1998, preprint, astro-ph/98\,03\,195.

\bibitem[\protect\citeauthoryear{Panagia}{1998}]{Panagia:98}
Panagia, N. 1998, preprint.

\bibitem[\protect\citeauthoryear{Parodi \& Tammann}{1998}]{Parodi:Tammann:98}
Parodi, B. \& Tammann, G.~A. 1998, to be published.

\bibitem[\protect\citeauthoryear{Perlmutter{\em
  ~et~al.}}{1998}]{Perlmutter:etal:98}
Perlmutter, S.,{\em ~et~al.} 1998, {\em Nature} {\bf 391},~51.

\bibitem[\protect\citeauthoryear{Pritchet \& van~den
  Bergh}{1987}]{Pritchet:vandenBergh:87}
Pritchet, C.~J. \& van~den Bergh, S. 1987, {\em ApJ} {\bf 318},~507.

\bibitem[\protect\citeauthoryear{Reid}{1998}]{Reid:98}
Reid, I.~N. 1998, {\em AJ} {\bf 115},~204.

\bibitem[\protect\citeauthoryear{Rephaeli}{1995}]{Rephaeli:95}
Rephaeli, Y. 1995, {\em ARA\&A} {\bf 33},~541.

\bibitem[\protect\citeauthoryear{Rephaeli \&
  Yankovitch}{1997}]{Rephaeli:Yankovitch:97}
Rephaeli, Y. \& Yankovitch, D. 1997, {\em ApJ} {\bf 481},~L55.

\bibitem[\protect\citeauthoryear{Saha{\em ~et~al.}}{1997}]{Saha:etal:97}
Saha, A., Sandage, A., Labhardt, L., Tammann, G.~A., Macchetto, F.~D., \&
  Panagia, N. 1997, {\em ApJ} {\bf 486},~1.

\bibitem[\protect\citeauthoryear{Saio \& Gautschy}{1998}]{Saio:Gautschy:98}
Saio, H. \& Gautschy, A. 1998, {\em ApJ}, in press.

\bibitem[\protect\citeauthoryear{Sandage}{1975}]{Sandage:75}
Sandage, A. 1975, {\em ApJ} {\bf 202},~563.

\bibitem[\protect\citeauthoryear{Sandage}{1993a}]{Sandage:93}
Sandage, A. 1993a, {\em ApJ} {\bf 106},~687 and 703.

\bibitem[\protect\citeauthoryear{Sandage}{1993b}]{Sandage:93a}
Sandage, A. 1993b, {\em ApJ} {\bf 402},~3.

\bibitem[\protect\citeauthoryear{Sandage}{1993c}]{Sandage:93b}
Sandage, A. 1993c, {\em ApJ} {\bf 404},~419.

\bibitem[\protect\citeauthoryear{Sandage}{1994}]{Sandage:94}
Sandage, A. 1994, {\em ApJ} {\bf 430},~13.

\bibitem[\protect\citeauthoryear{Sandage}{1996a}]{Sandage:96a}
Sandage, A. 1996a, {\em AJ} {\bf 111},~1.

\bibitem[\protect\citeauthoryear{Sandage}{1996b}]{Sandage:96b}
Sandage, A. 1996b, {\em AJ} {\bf 111},~18.

\bibitem[\protect\citeauthoryear{Sandage}{1998}]{Sandage:98a}
Sandage, A. 1998, private communication.

\bibitem[\protect\citeauthoryear{Sandage \& Bedke}{1988}]{Sandage:Bedke:88}
Sandage, A. \& Bedke, J. 1988, {\em Atlas of Galaxies useful to measure the
  Cosmological Distance Scale}, Washington: NASA.

\bibitem[\protect\citeauthoryear{Sandage{\em ~et~al.}}{1976}]{Sandage:etal:76}
Sandage, A., Kristian, J., \& Westphal, J.~A. 1976, {\em ApJ} {\bf 205},~688.

\bibitem[\protect\citeauthoryear{Sandage \&
  Tammann}{1971}]{Sandage:Tammann:71}
Sandage, A. \& Tammann, G.~A. 1971, {\em ApJ} {\bf 167},~293.

\bibitem[\protect\citeauthoryear{Sandage \&
  Tammann}{1982}]{Sandage:Tammann:82}
Sandage, A. \& Tammann, G.~A. 1982, {\em ApJ} {\bf 256},~339.

\bibitem[\protect\citeauthoryear{Sandage \&
  Tammann}{1988}]{Sandage:Tammann:88}
Sandage, A. \& Tammann, G.~A. 1988, {\em ApJ} {\bf 328},~1.

\bibitem[\protect\citeauthoryear{Sandage \&
  Tammann}{1993}]{Sandage:Tammann:93}
Sandage, A. \& Tammann, G.~A. 1993, {\em ApJ} {\bf 415},~1.

\bibitem[\protect\citeauthoryear{Sandage \&
  Tammann}{1995}]{Sandage:Tammann:95}
Sandage, A. \& Tammann, G.~A. 1995, {\em ApJ} {\bf 446},~1.

\bibitem[\protect\citeauthoryear{Sandage \&
  Tammann}{1998}]{Sandage:Tammann:98}
Sandage, A. \& Tammann, G.~A. 1998, {\em MNRAS} {\bf 293},~L23.

\bibitem[\protect\citeauthoryear{Sandage, Tammann, \&
Hardy}{1972}]{Sandage:etal:72}
Sandage, A., Tammann, G.~A., \& Hardy, E. 1972, {\em ApJ} {\bf 172},~253.

\bibitem[\protect\citeauthoryear{Schechter{\em
  ~et~al.}}{1997}]{Schechter:etal:97}
Schechter, P.~L.,{\em ~et~al.} 1997, {\em ApJ} {\bf 475},~L85.

\bibitem[\protect\citeauthoryear{Schmidt{\em ~et~al.}}{1994}]{Schmidt:etal:98}
Schmidt, B.~P.,{\em ~et~al.} 1994, {\em Science} {\bf 279},~1298.

\bibitem[\protect\citeauthoryear{Tammann}{1988}]{Tammann:88}
Tammann, G.~A. 1988, in: {\em The Extragalactic Distance Scale}, eds.
  S.~van~den Bergh, \& C.~J. Pritchet, San Francisco: ASP, p.~282.

\bibitem[\protect\citeauthoryear{Tammann}{1998}]{Tammann:98}
Tammann, G.~A. 1998, in: {\em Cosmological Parameters and Evolution of the
  Universe}, ed. K.~Sato, {\em I.\,A.\,U. Symp.}, {\bf 183}, in press.

\bibitem[\protect\citeauthoryear{Tammann \&
  Federspiel}{1997}]{Tammann:Federspiel:97}
Tammann, G.~A. \& Federspiel, M. 1997, in: {\em The Extragalactic Distance
  Scale}, eds. M.~Livio, M.~Donahue, \& N.~Panagia, Cambridge: Cambridge
  University Press, p.~137.

\bibitem[\protect\citeauthoryear{Tammann \&
  Sandage}{1995}]{Tammann:Sandage:95}
Tammann, G.~A. \& Sandage, A. 1995, {\em ApJ} {\bf 452},~16.

\bibitem[\protect\citeauthoryear{Tammann{\em ~et~al.}}{1997}]{Tammann:etal:97}
Tammann, G.~A., Sandage, A., Saha, A., Labhardt, L., Macchetto, F.~D., \&
  Panagia, N. 1997, in: {\em Thermonuclear Supernovae}, eds. P.~Ruiz-Lapuente,
  R.~Canal, \& J.~Isern, Dordrecht: Kluwer, p.~735.

\bibitem[\protect\citeauthoryear{Theureau{\em
  ~et~al.}}{1997}]{Theureau:etal:97}
Theureau, G.,{\em ~et~al.} 1997, {\em A\&A} {\bf 322},~730.

\bibitem[\protect\citeauthoryear{Thomsen{\em ~et~al.}}{1997}]{Thomsen:etal:97}
Thomsen, B., Baum, W.~A., Hammergren, M., \& Worthey, G. 1997, {\em ApJL} {\bf
  483},~37.

\bibitem[\protect\citeauthoryear{Tonry}{1997}]{Tonry:97}
Tonry, J.~L. 1997, in: {\em The Extragalactic Distance Scale}, eds. M.~Livio,
  M.~Donahue, \& N.~Panagia, Cambridge: Cambridge University Press, p.~297.

\bibitem[\protect\citeauthoryear{Webster{\em ~et~al.}}{1998}]{Webster:etal:98}
Webster, M., Hobson, M.~P., Lasenby, A.~N., Lahav, O., \& Rocha, G. 1998, {\em
  ApJL}, submitted.

\bibitem[\protect\citeauthoryear{Weedman}{1976}]{Weedman:76}
Weedman, D.~W. 1976, {\em ApJ} {\bf 203},~6.

\bibitem[\protect\citeauthoryear{Whitmore}{1997}]{Whitmore:97}
Whitmore, B.~C. 1997, in: {\em The Extragalactic Distance Scale}, eds.
  M.~Livio, M.~Donahue, \& N.~Panagia, Cambridge: Cambridge University Press,
  p.~254.

\end{thebibliography}

\end{document}